\documentclass[a4paper]{spie}  

\usepackage{amsmath,amsfonts,amssymb}
\usepackage[dvipdfmx]{graphicx}
\usepackage[colorlinks=true, allcolors=blue]{hyperref}

\newcommand{\um}{{\,$\mu$m}}

\newcommand{\fs}{FS}
\newcommand{\md}{MD}
\newcommand{\ct}{DTL}
\newcommand{\kt}{DifTL}

\newcommand{\ppi}{\,ppi}

\title{Spectrophotometric accuracy of spectra obtained from spectroscopic plates 
measured with a commercial ﬂatbed scanner}

\author[a]{Kenshi Yanagisawa}
\author[b]{Reiko Furusho}
\author[a]{Shiomi Nemoto}
\author[a]{Toshihiro Kasuga}
\author[a]{Yumi Iwashita}
\author[a]{Jun-ichi Watanabe}

\affil[a]{National Astronomical Observatory of Japan, 
Mitaka, Tokyo,  181-8588,  Japan}
\affil[b]{Tsuru University,  Tsuru, Yamanashi, 402-8555, Japan}

\authorinfo{Further author information: (Send correspondence to K.Y.)\\
K.Y.: E-mail: yanagisawa.kenshi@nao.ac.jp, Telephone: +81 422 34 3900 ex. 3530}

\begin{document} 

\maketitle

\begin{abstract}
The intensity spectra recovered from the spectroscopic photographic plates were compared with the CCD spectra, and we found the difference between the two was 2.5\%.
The measurements were taken using a commercial flatbed scanner instead of a microdensitometer.
The results indicate that the following two statements are inaccurate: 
(1) Correct measurement of spectroscopic plates requires a microdensitometer, and a commercial flatbed scanner is insufficient;
(2) The spectrophotometric accuracy of spectroscopic plate is approximately 10\% accurate.
These results will encourage the creation and publication of a digital archive of spectroscopic plates.
This paper presents the measurement of spectroscopic plates, the method of recovering intensity spectra from photographic density spectra, the results of comparison with CCD spectra, and discusses the causes of the high accuracy spectra obtained with a commercial flatbed scanner.
\end{abstract}


\section{Introduction}
\label{sec:intro}

The photographic plates were used over a century, from the late 1800s to the late 1900s. 
It is estimated that ten million plates have been accumulated worldwide \cite{2019AN....340..690H}. 
By utilizing these resources, it is possible to investigate the temporal variations of celestial bodies (transient, evolutionary, repetitive) over long periods. 
The International Astronomical Union has issued three recommendations in recognition of the importance of these irreplaceable research resources. 
The points of these recommendations is to share and make available the lists of plates and to digitize the plates and share the data.
In response to these recommendations, major research institutions have been digitizing the imaging plates, and some have been made publicly available as digital archives, being used for research and education \cite{1996AAS...188.5422C, 2009ASPC..410..101G, 2024arXiv240417355E}. 
On the other hand, no digital archive of spectral plates exists except for that of the Dominion Astrophysical Observatory\cite{Bohlender2012, Bohlender2012a}.
Although the digitization of spectroscopic plates seems feasible in a short time using readily available commercial flatbed scanners (hereafter referred to as \fs{}), 
several concerns have been raised regarding the use of \fs{}s for the digitization of plates. 
Incidentally, the Dominion Astrophysical Observatory's spectroscopic plate archive used a microdensitometer (hereafter referred to as \md{}) for plate digitization.


Common concerns about \fs{} include insufficient optical resolution\cite{2009ASPC..410..111S} and the inclusion of scattered light\cite{Griffin2015, griffin_balona_2017}.
The effective optical resolution of \fs{} is typically less than 2,400 dpi, which is not as high as the resolution of \md{}. 
With low optical resolution, it is not possible to extract all the information inherent in the plate.
In addition, because \fs{} inherently captures scattered light from the plate emulsion, the contrast of the transmitted image is lower and the accuracy of density measurement is inferior to \md{}, which does not capture scattered light. 
Concerns have also been raised that scattering can change the spectral shape when transmitted light is emitted from a position some distance from the incident position.
Therefore, there is an opinion that \md{} is best suited for measuring spectroscopic plates. 
However, there are very few \md{}s available worldwide today. 
If \fs{} cannot be used to digitize spectral plates, digital archiving of spectroscopic plates will not be realized in the future.

%
%

In this study, we aim to quantitatively evaluate the aforementioned concerns by digitizing spectroscopic plates using a \fs{} and comparing the recovered intensity spectra with CCD spectra. 
For digitization, we used a commercial \fs{} with the highest optical resolution currently available. 
The plates used are high-dispersion spectroscopic plates from the OAO. 
In this paper, we refer to the intensity spectra recovered from the plates as the {\bf Plate spectra}. 
The CCD spectra were obtained from the ELODIE archive \cite{Moultaka2004}. 
CCDs are superior to plates in terms of linearity, and their linearity does not depend on wavelength, making them an ideal reference.
We focused on the differences between the plate spectra and CCD spectra because all the concerns related to photographic plates and their measurement are reflected in these differences. 
If the differences are small, it can be judged that the impact of various concerns is also small.

The results of our trial were unexpected: the difference between the plate spectra and CCD spectra was 2.5\% on average. 
Previously, the accuracy of plate spectra was said to be 10\% \cite{acp-6-2231-2006}, but this study achieved an accuracy that is four times higher on average. 
This paper explains our trial and examines the reasons for the minimal differences between the plate spectra and CCD spectra.
This paper is structured as follows. Section \ref{sec:plate} introduces the spectroscopic plates processed. 
Section \ref{sec:digitize} describes the digitization of the spectroscopic plates, 
and Section \ref{sec:fukugen} explains the recovering processing of the plate spectra. 
In Section \ref{sec:comp}, we compare the plate spectra with the CCD spectra, 
and in Section \ref{sec:discuss}, we discuss the reasons for the small differences between them.

\section{spectroscopic plates}
\label{sec:plate}

\renewcommand{\arraystretch}{1.0}
\begin{table}
\centering
\caption{List of processed plates}
\label{tab:plate}
{\small
\begin{tabular}{cccccccccc}
\hline
Plate & Object & Sp. type & Date & Filter & Emulsion & Exposure & Developer &Grating & $\beta$ \\
 &  &  &  &  &  & (min.) &  & (gr./mm) & (deg.) \\
\hline
C4-4693 & $\gamma$ Ser & F6V & 1976-07-14 & none & IIaO+ & 26 & Pandol & 1,200 & 345 \\
C4-4701 & $\alpha$ Lyr & A0Va & 1976-07-15 & ND0.5 & IIaO+ & 5 & Pandol & 1,200 & 345 \\
C4-5033 & $\alpha$ Peg & B9III & 1977-08-02 & none & IIaO & 4 & Pandol & 1,200 & 345 \\
C4-5038 & $\alpha$ Dra & A0Va & 1977-08-03 & none & IIaO & 17 & Pandol & 1,200 & 345 \\
C4-5047 & $\theta$ Boo & F7V & 1977-08-04 & none & IIaO & 25 & Pandol & 1,200 & 345 \\
C4-5048 & $\sigma$ Boo & F4VkF2mF1 & 1977-08-04 & none & IIaO & 36 & Pandol & 1,200 & 345 \\
C4-5197 & $\eta$ UMa & B3V & 1977-12-27 & none & IIaO & 5 & Pandol & 1,200 & 345 \\
C4-5243 & 78 Vir & ApEuCrSi & 1978-05-20 & none & IIaO & 37 & Pandol & 1,200 & 345 \\
C4-5252 & 17 Com & A0p & 1978-05-22 & none & IIaO+ & 40 & Pandol & 1,200 & 345 \\
\hline
\end{tabular}
}
\end{table}
\renewcommand{\arraystretch}{1.0}

In this study, we used nine spectroscopic plates acquired with the coud\'{e} spectrograph attatched to the 188\,cm telescope 
at the Okayama Astrophysical Observatory (OAO)\footnote{On March 31, 2018, Okayama Astrophysical Observatory completed its mission and became the "Subaru Telescope Okayama Branch Office".}. 
The coud\'{e} spectrograph, manufactured by Hilger \& Watts, integrates a blazed diffraction grating with Schmidt cameras. 
It was the fifth such high-dispersion spectrograph in the world and started operations in 1961\cite{Hearnshaw2009}.
This spectrograph was used for observations for 30 years, during which 11,000 plates were obtained.
There are two types of Schmidt cameras (F/4 and F/10), and observers could choose the camera according to their purposes. 
All the plates processed in this study were obtained with the F/4 camera. 
The properties of the plates are shown in Table~\ref{tab:plate}, which shows that the filter, plate emulsion, grating groove density (Grating, in gr/mm), and diffraction angle ($\beta$) are all the same: the nine plates observed the same wavelength range ($\lambda\lambda$ 4,100--4,900 \,\AA)  with the same resiprocal dispersion ($\sim$10\,\AA /mm).
For development, Fujifilm's Pandol developer was used instead of Kodak's D19. 
The observation targets were B--F type stars, all of which are included in the ELODIE archive.
The reason for selecting these plates was that these plates had been measured by \md{} in the past, and we originally planned to compare \md{} measurements with \fs{} measurements. 
However, since a more accurate reference, CCD spectra, was found, no comparison with \md{} was made.

\section{Digitalization of plates}
\label{sec:digitize}

For digitizing the plates, we employed the RPS-4800 \fs{} from iMeasure Inc. 
This scanner offers the highest optical resolution among the currently available \fs{} models. 
The resolution of the RPS-4800 in the primary scanning direction was evaluated using the slant-edge method\cite{Burns2000}, resulting in a line width function FWHM of 11.2\um, equivalent to 2,270\,dpi.
Before scanning, the spectral plates were placed on the glass surface so that their dispersion direction was parallel to the primary scanning direction of the \fs{}. 
This setup was chosen to minimize image distortion\cite{Jones2012DimensionalMV, Wyatt:17}. 
In addition, the plate emulsion was in contact with the glass surface of the \fs{} to prevent interference fringes.

%

The scanning parameters were as follows.
The scanning resolution was set to 2,400\ppi{} (10.6\um/pix). 
Considering that the slit image width on the plates for the Okayama coud\'{e} spectrograph is 20\um, 2,400\ppi{} is a reasonable sampling pitch. 
The gamma value, which indicates the tonal response characteristic of the captured data, was set to 1.0. 
Digitization was performed at 16 bits, and the output was in monochrome, uncompressed TIFF format. 
With these settings, the relative transmittance value of the plates to the full range value (65,535) was obtained. 
Subsequently, the transmittance images were converted to density images and output in FITS format. 
Density is defined as the base-10 logarithm of the reciprocal of transmittance. 
The density in this paper adheres to this definition and is not relative to the fog level. 
It should be noted that the RPS-4800’s control software, iMeasure Scan, was designed with scientific measurement in mind.
Unlike some \fs{} control software that artificially enhances image resolution through software operations\cite{2014aspl.conf...15O, 2006ASPC..351..587D}, 
iMeasure Scan does not perform such enhancements.

\section{Processing of plates}
\label{sec:fukugen}

\begin{figure}
\centering
\includegraphics[scale=0.7]{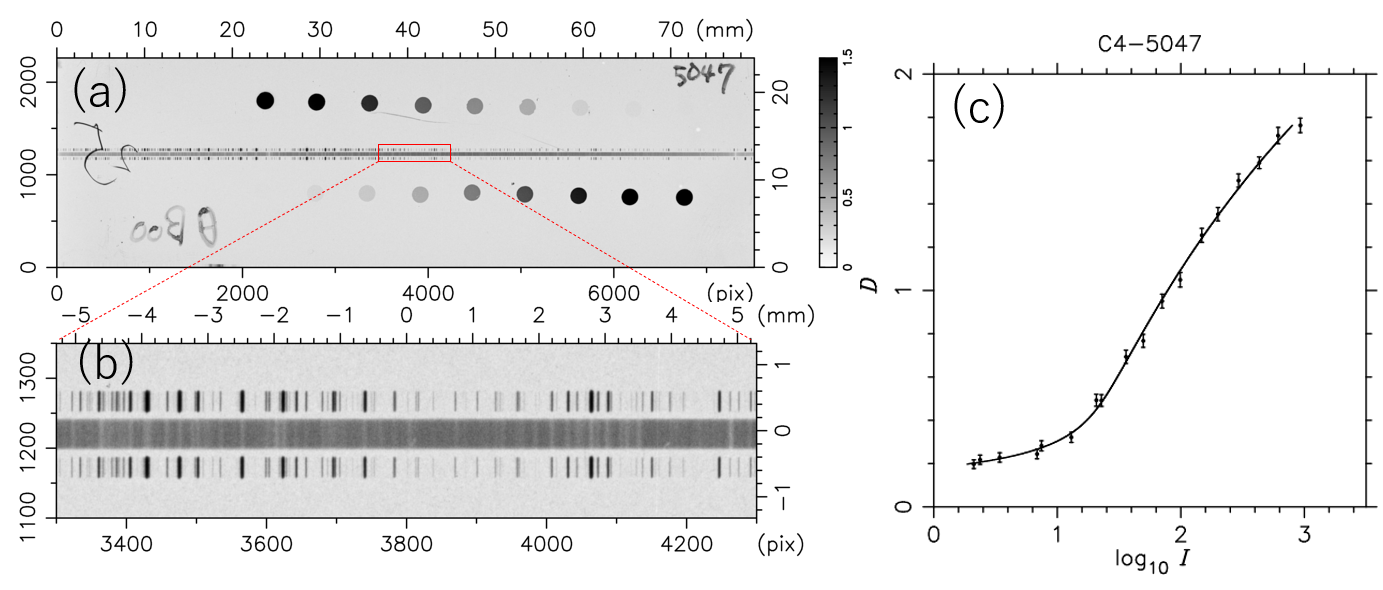}
\caption{
(a) An image showing the entire area (80\,mm$\times$24\,mm) of the spectroscopic plate, C4-5047, obtained with the coud\'{e} spectrograph at the Okayama Astrophysical Observatory. 
The stellar spectrum of $\theta$ Boo lies in the center, flanked by wavelength calibration spectra (iron neon). 
The circles ($\phi$=2.3\,mm) of different densities are spot sensitometer calibration elements relating  photographic density to exposure.
The unit of the scale bar is photographic density.
(b) An magnified image of the central part of the plate (10.5\,mm$\times$2.5\,mm). The stellar spectrum and the wavelength calibration spectra can be observed.
(c) The characteristic curve for plate C4-5047 showing the relation between photographic density ($D$), derived from the measurement of circles by the spot sensitometer, and exposure ($I$). 
The density range 0.3 to 1.8 is called the linear part.
}
\label{fig:plate}
\end{figure}

This section describes the processing method for extracting one-dimensional spectra of intensity scales from density images and the results of the processing.

\subsection{Methods}


Figure \ref{fig:plate} shows the density scale image and characteristic curve of plate C4-5047.
Figure~\ref{fig:plate}(a) depicts the entire plate, while Figure~\ref{fig:plate}(b) is an enlarged image of the red rectangle located in the center of (a). 
These images confirm the presence of three elements on the plate: the stellar spectrum, the wavelength calibration source spectra, and the calibration spots. 
The observed object is the F-type star $\theta$ Boo, which shows numerous absorption lines. 
The spectrum of the wavelength calibration source, which flanks the stellar spectrum, is used to correlate the positions on the plate with the wavelengths. 
The calibration spots are composed of 18 circles of $\phi$=2.3\,mm with varying densities, 16 of which can be observed in the Figure~\ref{fig:plate}(a). 
These circles are calibration elements from the spot sensitometer, and their purpose is to relate density to exposure.
Photographic photometry is a method of estimating the intensity of light by measuring the blackeness produced by the photographic emulsions's reaction to light. 
For this, a calibration light source with a known intensity ratio is required, and one form of this is the spot sensitometer.
Figure~\ref{fig:plate}(c) illustrates the characteristic curve or H-D curve\cite{https://doi.org/10.1002/jctb.5000090508}, which depicts the relation between the density and exposure of plate C4-5047.

%

The following processes are to be performed:
1. Create a characteristic curve, which is the relation between density and exposure, from the spots.
2. Apply the characteristic curve to convert the density image to an intensity image.
3. Subtract the background, extract the object spectrum image, and add them in the spatial direction to obtain the one-dimensional spectra of position and intensity.
4. From the wavelength calibration spectrum, relate the position on the plate to the wavelength, and convert the position and intensity to wavelength and intensity.

\subsection{Results}

\renewcommand{\arraystretch}{1.0}
\begin{table}
\centering
\caption{Summary of processed results}
\label{tab:results}
{\small
\begin{tabular}{ccccccccccc}
\hline
Plate & $\lambda\lambda$(\AA) & $\lambda_{c}$(\AA) & R(\AA/mm) & N & $\sigma_{\lambda}$(\AA) & $\lambda/\Delta\lambda$ & DR$_{\mathrm{TS}}$ & DR$_{\mathrm{SP}}$ & SNR & $\sigma_{R}$ \\
\hline
C4-4693 & 4,098--4,883 & 4,491 & 9.87 & 270 & 0.05 & 16,000 & 0.18--1.51 & 0.24--0.85 & 26   & 0.038\\
C4-4701 & 4,102--4,883 & 4,493 & 9.82 & 293 & 0.05 & 16,000 & 0.18--1.60 & 0.44--1.20 & 75   & 0.018\\
C4-5033 & 4,097--4,881 & 4,489 & 9.86 & 250 & 0.04 & 14,000 & 0.17--1.60 & 0.25--1.01 & 57   & 0.026\\
C4-5038 & 4,099--4,881 & 4,490 & 9.83 & 306 & 0.03 & 16,000 & 0.19--1.67 & 0.33--0.88 & 49   & 0.027\\
C4-5047 & 4,100--4,882 & 4,491 & 9.83 & 314 & 0.04 & 15,000 & 0.20--1.76 & 0.34--0.91 & 49   & 0.024\\
C4-5048 & 4,101--4,883 & 4,492 & 9.84 & 324 & 0.03 & 15,000 & 0.19--1.60 & 0.26--0.71 & 38   & 0.024\\
C4-5197 & 4,083--4,867 & 4,475 & 9.85 & 300 & 0.04 & 16,000 & 0.21--1.74 & 0.42--1.16 & 64   & 0.021\\
C4-5243 & 4,094--4,879 & 4,487 & 9.86 & 280 & 0.04 & 16,000 & 0.21--1.46 & 0.19--0.72 & 32   & 0.028\\
C4-5252 & 4,097--4,882 & 4,490 & 9.87 & 267 & 0.04 & 17,000 & 0.19--1.37 & 0.24--0.62 & 47   & 0.023\\
\hline
\end{tabular}
}
\end{table}
\renewcommand{\arraystretch}{1.0}

Table \ref{tab:results} shows the results of the recovery process. 
From this table, it can be seen that the wavelength coverage ($\lambda\lambda$), the central wavelength ($\lambda_{c}$), and the resiprocal dispersion (R) are almost the same for all plates.
This reflects the fact that the plates were obtained with the same spectrograph settings\footnote{The diffraction angle ($\beta$) setting of the diffraction grating is the same (see Section \ref{sec:plate}).}.

The number of emission lines (N) used for wavelength calibration was 290 on average, 
and the wavelength calibration accuracy ($\sigma_{\lambda}$) was 0.04\,\AA. 
A fifth-order polynomial was used to correlate the position on the plate with the wavelength, 
and the wavelength calibration accuracy was determined by the standard deviation of the residuals. 
The length on the plate corresponding to a wavelength of 0.04\,\AA{} is 4\um, which is less than one pixel\footnote{There 
is room for improvement in wavelength calibration accuracy. 
In this study, the peak pixel of the wavelength calibration emission line was correlated with the wavelength. 
If the emission line centroid is determined using the emission line shape as a weight, the wavelength calibration accuracy can be improved.
}.

The wavelength resolution ($\lambda/\Delta\lambda$) was 15,000 on average. 
Wavelength resolution was determined by converting the full width at half maximum of isolated wavelength calibration source spectra to wavelength width ($\Delta\lambda$) and then dividing the corresponding wavelength ($\lambda$) by these values, taking the median of the resulting set. 

DR$_{\mathrm{TS}}$ and DR$_{\mathrm{SP}}$ represent the density range from the 1\%ile to the 99\%ile. 
The former corresponds to spots, while the latter pertains to stellar spectra. 
The calibration of the stellar spectra from the density to the intensity was performed appropriately, 
as indicated by DR$_{\mathrm{SP}}$ falling within the range of DR$_{\mathrm{TS}}$ for all plates.
Furthermore, the maximum value of DR$_{\mathrm{SP}}$ is located near the central part of the linear region, where the DQE \cite{1978mtap.conf..153F} is at its maximum. 

The SNR is the signal-to-noise ratio of the derived one-dimensional stellar spectrum. 
The average SNR of the nine plates was 50. 
The algorithm used to calculate the SNR was DER\_SNR \cite{Stoehr2008}.

\section{Spectrophotometric accuracy}
\label{sec:comp}

\begin{figure}
\centering
\vspace{2mm}
\begin{tabular}{cc}
\begin{minipage}[t]{0.48\linewidth}
\includegraphics[scale=0.65]{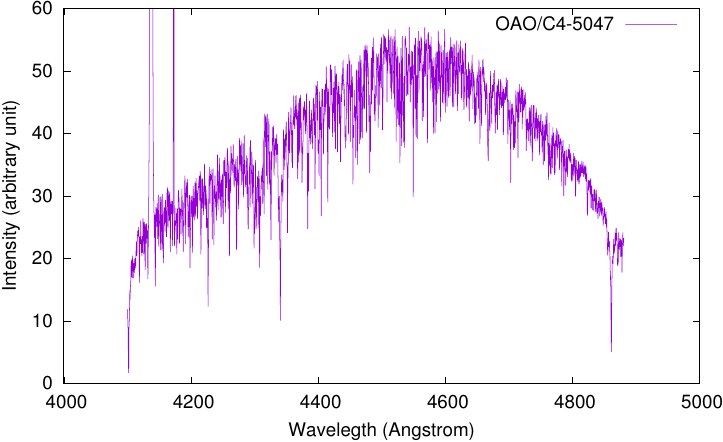}
\caption{Plate spectrum recovered from the spectoscopic plate, C4-5047, which observed $\theta$ Boo. The plate was obtained by the coud\'{e} spectrograph attached to the 188\,cm telescope at Okayama Astrophysical Observatory.}
\label{fig:oao}
\end{minipage}
&
\begin{minipage}[t]{0.48\linewidth}
\includegraphics[scale=0.65]{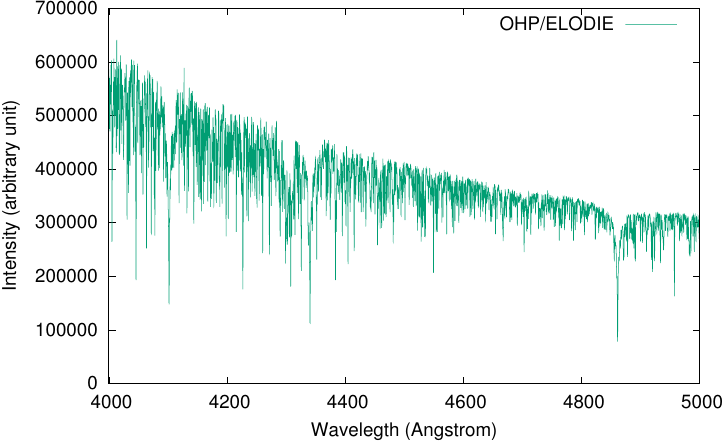}
\caption{CCD spectrum of $\theta$ Boo obtained by ELODIE, Haute-Provence Observatory.}
\label{fig:ohp}
\end{minipage}
\\
\end{tabular}
\end{figure}


In this section, we compare the spectra obtained from photographic plates with those from CCDs. 
Among the nine plates analyzed, we show C4-5047 as a representative, which observed the star $\theta$ Boo (F7V, B=4.56$^{m}$). 
Figure~\ref{fig:oao} shows the intensity spectrum recovered from the spectroscopic plate, C4-5047, while Figure~\ref{fig:ohp} shows the CCD spectrum of the same object downloaded from the ELODIE library. 
These figures indicate that the continuum level profiles of both spectra differ, with the plate spectrum showoing a convex shape. 
This difference directly reflects the spectroscopic sensitivity characteristics of the photographic plate and the efficiency of the spectrograph.



\renewcommand{\arraystretch}{1.0}
\begin{table}
\centering
\vspace{3mm}
\caption{ Characteristics of two spectra}
\label{tab:comp}
{\small
\begin{tabular}{p{2.5cm}p{2.5cm}p{2.5cm}}
\hline
 & OAO Plate & ELODIE CCD \\
 \hline
Object & \multicolumn{2}{c}{$\theta$ Boo} \\
Sp. type & \multicolumn{2}{c}{F7V} \\
Magnitude & \multicolumn{2}{c}{B=4.56} \\
Source & C4-5047 & elodie\_19990601\_0010.fits \\
Date & Aug. 4, 1977 & Jun. 1, 1999 \\
$\lambda\lambda$ (\AA) & 4,100--4,882 & 4,000--5,000 \\
$\Delta\lambda$(\AA) & 0.295 $\pm$ 0.056 & 0.132 \\
$\lambda/\Delta\lambda$ & 14,900 & 34,100 \\
Sampling(\AA/pix) & 0.104 & 0.05 \\
\hline
\end{tabular}
}
\end{table}
\renewcommand{\arraystretch}{1.0}

Characteristics for the plate and CCD spectra are shown in Table~\ref{tab:comp}. 
To directly compare these spectra, we corrected for differences in wavelength resolution and radial velocity, and normalized them at the continuum level. 
The wavelength resolution of the CCD spectrum is more than twice that of the plate spectrum. 
Therefore, the CCD spectra were convolved with a Gaussian function to match the wavelength resolution of the plate spectra. 
The radial velocity difference was corrected by adjusting the plate spectrum with a constant factor in the wavelength direction to match the CCD spectrum. 
Finally, both spectra were normalized to the continuum level.


\begin{figure}
\centering
\begin{tabular}{c}
\begin{minipage}[t]{0.95\linewidth}
\centering
\includegraphics[scale=0.7]{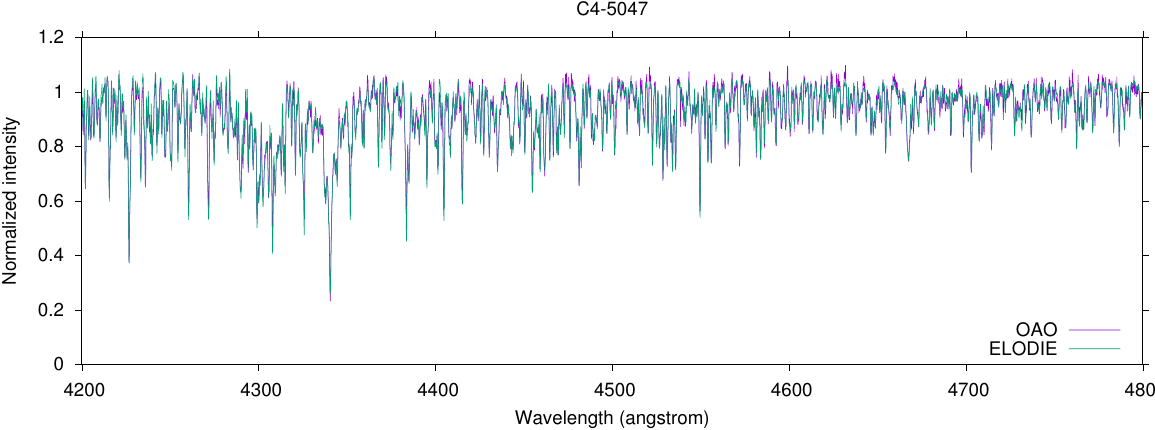}
\end{minipage}
\\
\begin{minipage}[t]{0.95\linewidth}
\centering
\includegraphics[scale=0.7]{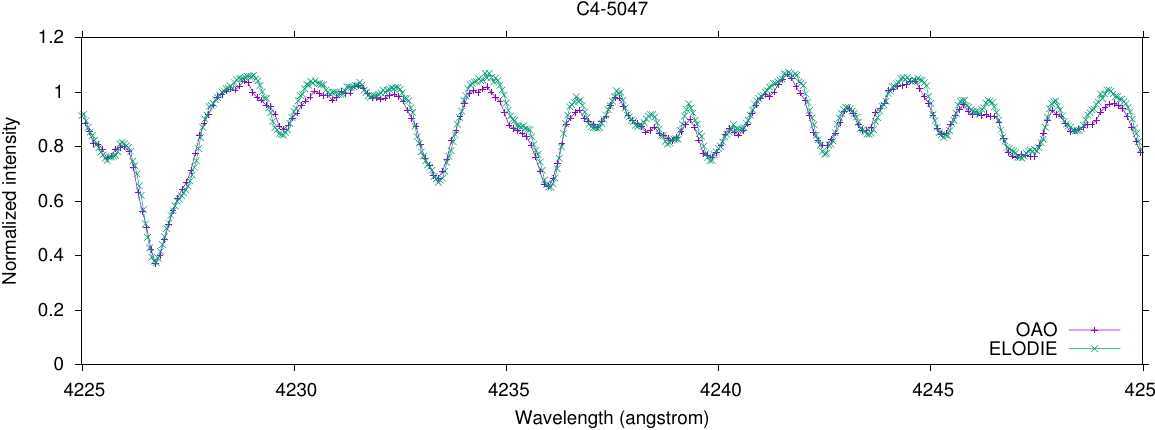}
\end{minipage}
\\
\begin{minipage}[t]{0.95\linewidth}
\centering
\includegraphics[scale=0.7]{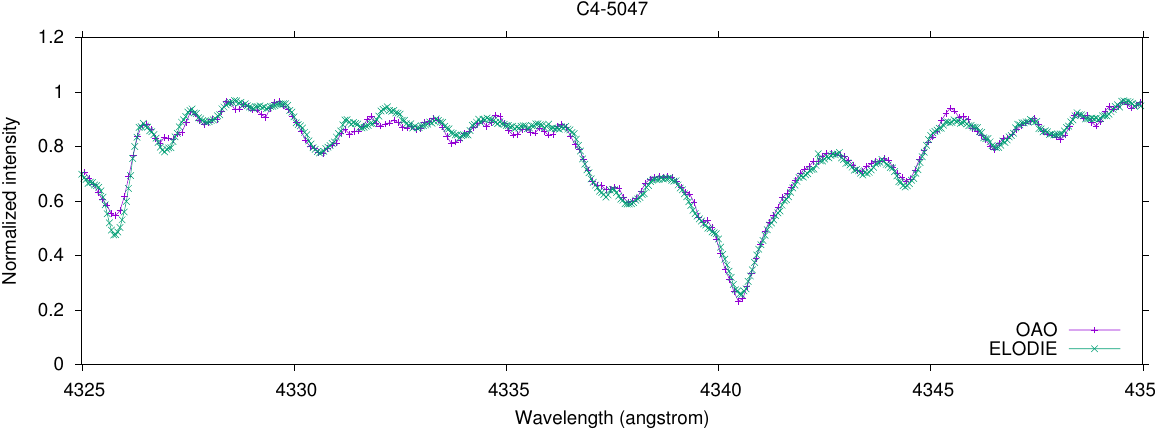}
\end{minipage}
\\\begin{minipage}[t]{0.95\linewidth}
\centering
\includegraphics[scale=0.7]{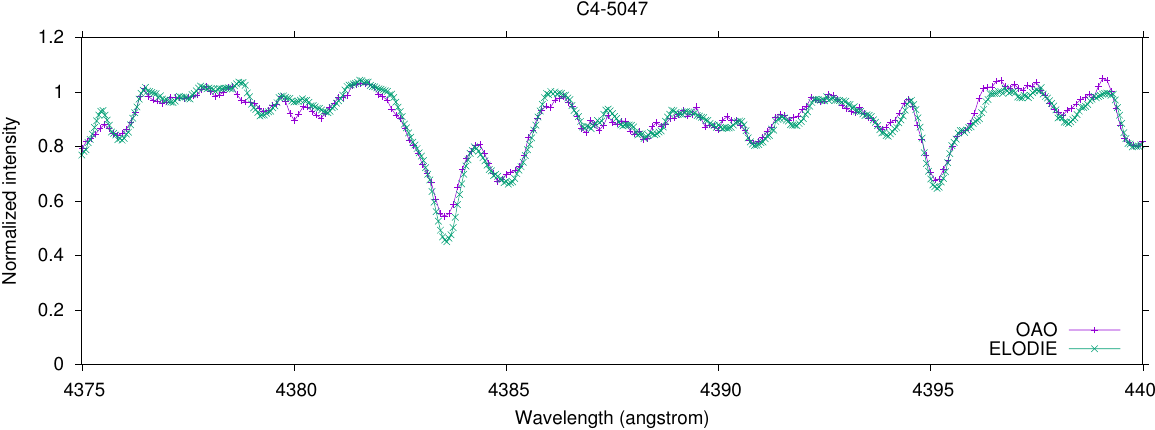}
\end{minipage}
\\
\end{tabular}
\caption{
Superimposed plate spectrum restored from C4-5047 and CCD spectrum after normalization.
Purple is the plate spectrum and green is the CCD spectrum. 
The top panel shows the wavelength range $\lambda\lambda$ 4,200--4,800 \AA. The bottom three panels selectively show wavelength regions containing deep absorption lines, each covering 25\,\AA. 
The H$_{\gamma}$ absorption line is visible in the right center of the second panel from the bottom.
}
\label{fig:comp1}
\end{figure}

Figure \ref{fig:comp1} superimposes the normalized spectra, consisting of four panels. 
The top panel has the widest wavelength coverage, spanning $\lambda\lambda$ 4,200--4,800\,\AA. 
The bottom three panels selectively display wavelength regions containing deep absorption lines, each covering 25\,\AA. 
In each panel, the purple line represents the spectroscopic plate spectrum, while the green line represents the CCD spectrum. 

The top panel shows almost no difference between the plate spectrum and the CCD spectrum over the entire wavelength range. 
The lower three panels, which provide enlarged views, show that the shapes of the absorption lines are nearly identical. 
Differences of up to about 9\% are observed at the bottoms of some deep absorption lines. 
However, since many other equally deep absorption lines match well, this discrepancy may be attributed to phenomena occurring during different observation periods.

\begin{figure}
\centering
\includegraphics[scale=0.8]{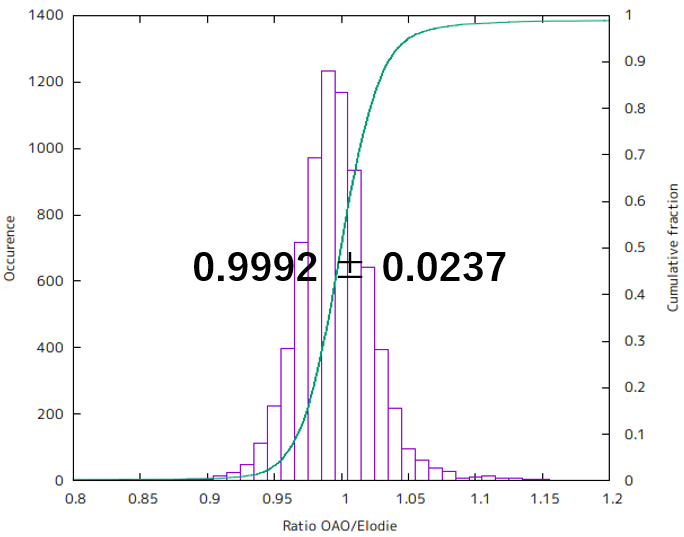}
\caption{
Histogram of the ratio of each wavelength obtained by dividing the plate spectrum from C4-5047 by the CCD spectrum. 
The distribution of the histogram is Gaussian with a standard deviation of 2.4\%; 
since the CCD spectrum is a reliable reference, a standard deviation of 2.4\% can be considered as the spectrophotometric accuracy of the spectroscopic plate. 
}
\label{fig:hist0}
\end{figure}


To quantify the spectral differences, we calculated the ratio by dividing the plate spectrum by the CCD spectrum for each object. 
Figure~\ref{fig:hist0} shows a histogram of the distribution of these ratios for the F-type star $\theta$ Boo. 
Despite the large number of absorption lines in the F-type star, the histogram shows a Gaussian and symmetric distribution. 
The mean of the ratios is 0.999, and the standard deviation is 2.4\%.

The standard deviations obtained similarly for the other eight plates are listed as $\sigma_{R}$ in the 11$^{th}$ column of Table~\ref{tab:results}. 
The maximum value of $\sigma_{R}$ is 3.8\%, and the average is 2.5\%. 
For F-type stars with numerous absorption lines, the average $\sigma_{R}$ increases slightly to 2.9\%. 
When converted to signal-to-noise ratio (SNR), the average value of $\sigma_{R}$ is 40, which is in good agreement with the average SNR value of 50 reported in Table~\ref{tab:results} (Section~\ref{sec:fukugen}).

\section{Discussion}
\label{sec:discuss}

Nine coud\'{e} spectroscopic plates obtained at OAO were digitized by a RPS-4800 \fs{}, 
and the intensity-scale plate spectra obtained by standard procedures were compared with the CCD spectra. 
The difference between the two spectra of the same source averaged 2.5\%, which was smaller than previously expected. 
This difference reflects all the concerns associated with photographic photometry, making individual separations challenging. 
This section discusses the reasons for these results.

\subsection{\fs{}s pick up scatterd light but allow photometry}

One of the concerns regarding \fs{}s is the inability to avoid scattered light generated by the plate emulsion.
Unfortunately, there are no reports on the transmission scattering characteristics of developed photographic plates. 
Therefore, in this subsection, we will summarize the general properties of direct transmitting light and diffuse transmitting light, 
predict the differences in characteristic curves between \fs{}s and \md{}s, 
and explain why \fs{}s could correctly convert density to intensity.


The light transmitted by a sample is typically divided into two components: direct transmitted light (hereafter referred to as  \ct{}) and diffuse transmitted light (hereafter referred to as \kt{}). 
In the case of photographic plates, the light reaching the \fs{} detector consists of \ct{} associated with vertically incident light and \kt{} associated with obliquely incident light. 
The \md{} measures only \ct{} because it does not illuminate the sample obliquely. 
Consequently, the \fs{} detector detects a greater quantity of transmitted light than the \md{} detector, with the exception of instances where the transmittance of the sample is equal to one. 
Furthermore, the ratio of \ct{} to \kt{} varies in accordance with the transmittance of the sample. 
When the transmittance is nearly equal to one, the difference between the \ct{} and \kt{} is minimal. 
However, when the transmittance approaches zero, the difference between the \ct{} and \kt{} is greater. 
Since \md{} measures only the \ct{}, it is able to measure low transmittance regions with a higher signal-to-noise ratio than \fs{}.


If the same plate is measured separately for \md{} and \fs{} and their characteristic curves are plotted on the same graph, it can be observed that the characteristic curve for \fs{} is consistently positioned below that of \md{}. 
As the exposure increases (moving the horizontal axis to the right), the vertical distance between the characteristic curves increases. 
And at the right end of the characteristic curve, \fs{} cannot distinguish the difference in blackness, and the slope of the characteristic curve may be smaller.
 However, the difference in the amount of scattered light has no effect on the characteristic curve other than the aforementioned: 
 the \fs{} outputs a density that includes \kt{}, but as long as the same density is output for the same "blackness," the "blackness" can be converted to intensity via the characteristic curve of the \fs{}.

The linear part of the characteristic curve represents the region where the density and the normal logarithm of the exposure are approximately proportional. 
This region is of significant importance, as it determines the dynamic range of photographic photometry. 
Figure \ref{fig:plate}(c) shows the characteristic curve obtained by digitizing the high-dispersion spectroscopic plate C4-5047 using a \fs{}. 
This figure clearly shows the linear part.
The density range of the stellar spectrum is 0.34–0.91 (DR$_{\mathrm{SP}}$ in Table \ref{tab:results}), which falls within the density range of the linear part (0.3–1.8). 
The density range of the stellar spectra for all the spectroscopic plates digitized in this study was on average 0.3–0.9. 
Because the same plate emulsion and development method were used, the shape of the characteristic curve did not vary significantly. 
Therefore, nearly all the samples in this study could be ideally converted from density to intensity.

\subsection{\fs{}s have an optics that is less likely to pick up scattered light}


The reason why the characteristic curve of \fs{} does not seem to be different from that of \md{} may have something to do with the fact that \fs{} optics are more difficult to pick up scattered light than \md{} optics.
In general, the aperture ratio of \fs{} optics is large; in the case of RPS-4800, it is F/9.5. 
The maximum solid angle from the sample to the \fs{} optics is 0.0087 sr, and the ratio of this solid angle to the solid angle of the hemisphere is 0.0014.
\kt{} is emitted in all directions, but only a small fraction of the total \kt{} reaches the \fs{} detector.
In contrast, the aperture ratio of the \md{} optics is small, about F/2.4.
The solid angle at which the sample looks into the \md{} optics is 0.13 sr.
This indicates that the \md{} optics is 15 times more likely to pick up \kt{} than the \fs{} RPS-4800.

\subsection{High-dispersion spectroscopic plates are less sensitive to \kt{}}

Now, the fact that the density of the processed plates was in the lower half of the linear part implies that the average transmittance of the stellar spectra was high. 
In general, in high dispersion spectroscopic plates, the illumination on the detector is not very high, so the density does not become large. 
Therefore, it can be said that high dispersion spectroscopic plates are less susceptible to the effect of \kt{}.
As mentioned above, when the transmittance is high, \kt{} is small compared to \ct{}, which means that from the perspective of SNR, \fs{} is not significantly disadvantaged compared to \md{}.


The OAO coud\'{e} spectroscopic plates tend to have a low density, which may make them somewhat advantageous for photometric measurements using a \fs{}. 
The nine photographic plates digitized with an \fs{} in this study were all developed with Pandol developer from Fuji Photo Film Co., Ltd. 
Compared to Kodak's D19 developer, Pandol produces a softer finish\cite{Hamajima1981}. 
In general, softer processing suppresses development effects such as the Eberhard effect\cite{KennethMees:31}, which is advantageous for accurately recording the shapes of absorption lines. 
In addition to being softer, Pandol is not only softer but also excels in graininess, providing a DQE equivalent to that of D19\cite{Hamajima1981}.

\subsection{Low distortion of absorption line profiles due to scattering}

If scattered light on the plate emulsion is transmitted from locations far from the incident position, the absorption line profiles of the spectroscopic plate may be distorted. 
We have aligned the dispersion direction of the spectroscopic plate parallel to the primary scanning direction of the \fs{}, making it a setup where this problem is more likely to manifest. 
However, in this study, we found that the distortion was either not present or too small to be detected.

As shown in Figure~\ref{fig:comp1}, the plate spectrum and the CCD spectrum are in good agreement.
We succeeded in obtaining images that nearly preserve the resolution of the photographic plates\footnote{We captured the 20\um{} slit image on the plate using a \fs{} with an optical resolution of 11.2\um.}. 
Each pixel of the captured image corresponds to the purple crosses in Figure \ref{fig:comp1}. 
In this figure, the absorption lines of the F-type star are sampled with a sufficient number of pixels. 
This suggests that the scattering at the emulsion is not large enough to distort the broad absorption lines of the F-type stars. 
In plates with high photographic density, which are more scattering, and spectra with narrow absorption lines
—such as those from bright, low-temperature stars—using a \fs{} might reveal distortion in the absorption line profiles.

\subsection{Cause of no systemic difference in wavelength direction}
%

In the top spectral comparison diagram of Figure~\ref{fig:comp1}, the absence of systematic differences in the wavelength direction is due to the minimal wavelength dependence of the characteristic curve. 
The OAO coud\'{e} spectrograph was equipped with a spot sensitometer to calibrate density to intensity. 
This spot sensitometer does not provide wavelength-specific characteristic curves.

If the wavelength dependence of the characteristic curve were significant, 
applying the characteristic curve derived from spots  across all wavelength regions would introduce systematic errors into the spectra. 
The nine plates processed in this study all used the IIa-O emulsion, covering the wavelength range of 4,100--4,900 \,\AA. 
In this wavelength range, the characteristic curve shape of the IIa-O emulsion is almost identical \cite{1986ASSL..125..209H, 1966PASP...78..537K}. 
Thus, the wavelength variation of the input-output characteristics is minimized, allowing the spot sensitometer to appropriately calibrate density to intensity.

\section{Conclusions}


In this study, nine high dispersion spectroscopic plates obtained at Okayama Astrophysical Observatory were digitized using the RPS-4800 flatbed scanner from iMeasure Inc. and each was compared to a CCD spectrum from ELODIE archive.
Comparisons between CCD spectra and plate spectra have rarely been made, and this is the first comparison using high dispersion plate spectra digitized with the flatbed scanner.
The two spectra agreed well, with an average difference of 2.5\%, which we concluded to be the spectrophotometric accuracy.
This value is four times better than the previously considered spectrophotometric accuracy of 10\%.


The differences between plate and CCD spectra reflect several concerns related to photographic photometry. 
These concerns include the effect of scattered light in the plate emulsion, development effects, the wavelength dependence of the characteristic curve, etc. 
In particular, the flatbed scanner was initially thought to have disadvantages over the microdensitometer because of its nature to capture scattered light. 
While it is true that the microdensitometers have superior optical resolution and can measure density more accurately than the flatbed scanners, it is undeniable that a high spectrophotometric accuracy of 2.5\% was achieved despite the shortcomings of the flatbed scanners. 
This finding suggests that the flatbed scanner is sufficiently useful for digitizing spectroscopic plates. 
The RPS-4800 flatbed scanner has the highest optical resolution among current products. 
This ability to measure without significant loss of information from the plates contributed greatly to the results of this study.


\acknowledgments

We received considerable assistance from many individuals in conducting this study. 
We are grateful to Kokusai Microfilm Co., Ltd. for providing the opportunity to digitize the spectroscopic plates. 
Without the ability to digitize using a high-precision flatbed scanner, this study would not have been possible. 
We also thank iMeasure Inc. for providing information about the optical system of the RPS-4800 flatbed scanner and for their support with inquiries and consultations related to the measurement of photographic plates.

\bibliography{plate_spie_2024} 
\bibliographystyle{spiebib} 

\end{document}